\documentclass[conference]{IEEEtran}
\IEEEoverridecommandlockouts
\usepackage{cite}
\usepackage{amsmath,amssymb,amsfonts}
\usepackage{algorithmic}
\usepackage{graphicx}
\usepackage{textcomp}
\usepackage{xcolor}

\graphicspath{{fig/}}
\usepackage{mathtools}
\usepackage{subfig}
\usepackage{siunitx}
\usepackage{bmpsize}
\usepackage[abs]{overpic}
\usepackage{adjustbox}
\usepackage[labelsep=period]{caption}
\renewcommand{\thetable}{\Roman{table}}
\def\tablename{TABLE}
\usepackage{booktabs,array}
\usepackage{multirow}
\usepackage{balance}

\renewcommand{\baselinestretch}{0.99} 

\usepackage{pgfplots}
\usepackage{grffile}
\pgfplotsset{compat=newest}
\usetikzlibrary{plotmarks}
\usetikzlibrary{arrows.meta}
\usepgfplotslibrary{patchplots}
\usepackage{colortbl}%

\usepackage[colorlinks=true,linkcolor=black, citecolor=black, urlcolor=black]{hyperref} 
\usepackage{bm}
\usepackage{upgreek}
\usepackage{units}

\newcommand{\midsepremove}{\aboverulesep = 0mm \belowrulesep = 0mm}
\midsepremove
\newcommand{\midsepdefault}{\aboverulesep = 0.605mm \belowrulesep = 0.984mm}
\midsepdefault

\definecolor{mycolor1}{rgb}{0.00000,0.44700,0.74100}%
\definecolor{mycolor3}{rgb}{0.92900,0.69400,0.12500}%
\definecolor{mycolor4}{rgb}{0.46667,0.67451,0.18824}%
\definecolor{mycolor2}{rgb}{0.85000,0.32500,0.09800}%

\definecolor{mycolor0}{rgb}{1,1,1}%
\definecolor{mycolor1}{rgb}{0.00000,0.45000,0.74000}%
\definecolor{mycolor2}{rgb}{0.85000,0.30000,0.10000}%
\definecolor{mycolor3}{rgb}{0.04706,0.47451,0.58039}%
\definecolor{mycolor4}{rgb}{0.00000,0.44700,0.74100}%
\definecolor{mycolor5}{rgb}{0.85000,0.32500,0.09800}%
\definecolor{mycolor6}{rgb}{0.00000,0.44700,0.74100}%
\definecolor{mycolor7}{rgb}{0.85000,0.32500,0.09800}%
\definecolor{mycolor8}{rgb}{0.92900,0.69400,0.12500}%
\definecolor{mycolor9}{rgb}{0.93000,0.69000,0.13000}%
\definecolor{mycolor10}{rgb}{0.49000,0.18000,0.56000}%
\definecolor{mycolor11}{rgb}{0.32000,0.49000,0.08000}%
\definecolor{mycolor12}{rgb}{0.85000,0.33000,0.10000}%
\definecolor{mycolor13}{rgb}{0.30000,0.42000,0.13000}%
\definecolor{mycolor14}{rgb}{0.64000,0.08000,0.18000}%
\definecolor{mycolor15}{rgb}{0.46667,0.67451,0.18824}%
\definecolor{mycolor16}{rgb}{0.49412,0.18431,0.55686}%

\newcommand{\myarrowdotted}[1][0.1pt]
{   \begin{tikzpicture}[overlay]
	\draw [->,>=stealth,line width=0.4mm,dashed,black] (-0.1, 0.1) -- (0.4, 0.1);
	\end{tikzpicture}
}

\newcommand{\myarrowsolid}[1][0.1pt]
{   \begin{tikzpicture}[overlay]
	\draw [->,>=stealth,line width=0.4mm,solid,black] (-0.1, 0.1) -- (0.4, 0.1);
	\end{tikzpicture}
}

\definecolor{blueelipse}{rgb}{0,0.2745,1}%
\definecolor{redelipse}{rgb}{0.784,0,0.1254}%
\definecolor{purpleelipse}{rgb}{0.5921,0.0470,0.8745}%
\newcommand{\myelipseblue}[1][0.1pt]
{   \begin{tikzpicture}[overlay]
	\draw [line width=0.5mm,blueelipse](0, 0.1) ellipse (1mm and 1mm);
	\end{tikzpicture}
}
\newcommand{\myelipsered}[1][0.1pt]
{   \begin{tikzpicture}[overlay]
	\draw [line width=0.5mm,redelipse](0, 0.1) ellipse (1mm and 1mm);
	\end{tikzpicture}
}
\newcommand{\myelipsepurple}[1][0.1pt]
{   \begin{tikzpicture}[overlay]
	\draw [line width=0.5mm,purpleelipse](0, 0.1) ellipse (1mm and 1mm);
	\end{tikzpicture}
}

\makeatletter
\renewcommand\footnotesize{%
	\@setfontsize\footnotesize\@ixpt{11.4}%
	\abovedisplayskip 8\p@ \@plus2\p@ \@minus4\p@
	\abovedisplayshortskip \z@ \@plus\p@
	\belowdisplayshortskip 4\p@ \@plus2\p@ \@minus2\p@
	\def\@listi{\leftmargin\leftmargini
		\topsep 4\p@ \@plus2\p@ \@minus2\p@
		\parsep 2\p@ \@plus\p@ \@minus\p@
		\itemsep \parsep}%
	\belowdisplayskip \abovedisplayskip
}
\makeatother

\def\BibTeX{{\rm B\kern-.05em{\sc i\kern-.025em b}\kern-.08em
    T\kern-.1667em\lower.7ex\hbox{E}\kern-.125emX}}
\begin{document}

\title{A MIMO Radar-based Few-Shot Learning Approach for Human-ID\\
}

\author{Pascal Weller\textsuperscript{~1},
	Fady Aziz\textsuperscript{~1},
	Sherif Abdulatif\textsuperscript{~2}, 
	Urs Schneider\textsuperscript{~1},
	Marco F. Huber\textsuperscript{~3,4}\\
\IEEEauthorblockA{\textsuperscript{1}Department~of~Bio-mechatronic~Systems,~Fraunhofer~IPA,~Stuttgart,~Germany\\
	\textsuperscript{2}Institute~of~Signal~Processing~and~System~Theory,~University~of~Stuttgart,~Germany\\
	\textsuperscript{3}Institute~for~Industrial~Manufacturing~and~Factory~Operation~IFF,~University~of~Stuttgart,~Germany\\
	\textsuperscript{4}Center~for~Cyber~Cognitive~Intelligence,~Fraunhofer~IPA,~Stuttgart,~Germany
	\thanks{Email: fady.aziz@ipa.fraunhofer.de}
	\thanks{\textbf{The first two authors contributed to this work equally.}}
}}



\maketitle

\begin{abstract}
	\par Radar for deep learning-based human identification has become a research area of increasing interest. It has been shown that micro-Doppler (\(\upmu\)-D) can reflect the walking behavior, through capturing the periodic limbs micro-motions. One of the main aspects is maximizing the number of included classes, while considering the real-time and training dataset size constraints. In this paper, a multiple-input-multiple-output (MIMO) radar is used to formulate micro-motion spectrograms of the elevation angular velocity (\(\upmu\)-\(\omega\)). The effectiveness of concatenating this newly-formulated spectrogram with the commonly used \(\upmu\)-D ones is investigated. To accommodate for non-constrained real walking motion, an adaptive cycle segmentation framework is utilized and a metric learning network is trained on half gait cycles (\(\mathbf{\approx}\)\unit[0.5]{s}). Studies on the effects of various numbers of classes (5--20), different dataset sizes, and varying observation time windows (\unit[1--2]{s}) are conducted. A non-constrained walking dataset of 22 subjects is collected with different aspect angles with respect to the radar. The proposed few-shot learning (FSL) approach achieves a classification error of \unit[11.3]{\%} with only \unit[2]{min} of training data per subject.	
\end{abstract}


\section{Introduction}
	\label{sec:intro}
	\par Person identification (human-ID) is of great interest for applications like surveillance, access control, and safety systems, where biometric characteristics are commonly used \cite{yang2020mu}. One of them is the radar-based micro-Doppler (\(\upmu\)-D) signature of human gait, which has recently been heavily investigated for human-ID \cite{pegoraro2020multiperson,zhao2019mid,abdulatif2019person}. The superposition of micro-motions causes such signatures during movement, e.g., swinging legs and arms \cite{chen2006micro,chen2019micro}. In contrast to common biometrics, remote sensing has the advantage of being non-cooperative, works for gait recognition on long distances, and is hard to disguise \cite{kose2013vulnerability, erdogmus2014spoofing, hadid2015biometrics}. Radar beats other visionary systems as it is privacy-conservative and can handle nearly any environmental condition e.g., dust, smoke, darkness, and sunlight \cite{radarSherif}.

	\par Multiple-input-multiple-output (MIMO) radar modules have been deployed to the task of human-ID. The interferometric analysis between different receiving antennas is used to estimate the target's angle of arrival (AoA) \cite{milligan2005modern}, where a tracking feature is added to \(\upmu\)-D-based human-ID applications \cite{pegoraro2020multiperson,zhao2019mid}. However, the rate of change of the AoA profiles can be monitored through time to reflect the micro-motions of the angular velocity. To the best of our knowledge, this wasn't previously investigated as an extra feature for enhancing the human-ID task. Such technique is different from the one presented in \cite{nanzer2010millimeter}, where the tangential angular velocity was derived independently of the AoA due to hardware limitations. The main aim was to extend the field of view (FOV) to include tangential scenarios.

	\par Deep learning (DL) has successfully been adapted to the topic of radar-based human-ID and has led to leaps in classification accuracy \cite{cao2018radar}. The majority of present research work is based on \(\upmu\)-D spectrograms and DCNN architectures \cite{papanastasiou2021deep, qiao2020human, vandersmissen2018indoor}. The benefit of more sophisticated DCNNs has been studied in \cite{yang2019person, jalalvand2019radar, addabbo2020gait}, while an explicit use of the range-Doppler maps is presented in \cite{yang2020mu, ni2020human}. The latter proves the effectiveness of transfer learning on the topic of radar-based human-ID. Thus, transfer learning is presented as a remedy for the requirement of multitudes of data for training DL approaches that could also be utilized for activity recognition \cite{gurbuz2019radar}. For the same purpose, few-shot learning (FSL) is deployed for human-ID in \cite{niazi2021radar, ni2021open}. FSL has the advantage of relying on less labeled samples to successfully train a DCNN \cite{wang2020generalizing} and performs better on unseen data. However, using \(\upmu\)-D signatures comes at the expense of the observation window and the number of classes.	
	
	\par  All mentioned papers solved the radar-based human-ID task based on an observation window of at least \unit[1]{s} of \(\upmu\)-D spectrogram data (one full gait cycle), with the majority being in the \unit[3]{s} range. Relying on longer observation windows enables the inclusion of more classes and identification of a less-constrained walking style. The work presented in \cite{abdulatif2019person} has accomplished classification on half gait cycles (\(\approx\)\unit[0.5]{s}). However, the study was conducted on restricted treadmill motion. To sum up, there is a trade-off between the real-time analysis and radar features as well as non-restricted walking and the number of classes. Moreover, relying on less training data is recommended as the human-ID application always requires the inclusion of new classes. Thus, walking for long time periods and retraining of the entire network is not applicable in real life.  	
	
	\par In this paper a human-ID task is considered, where 22 subjects walk according to their natural gait in a non-constrained manner. The framework is based on a MIMO radar to capture  \(\upmu\)-\(\omega\) and \(\upmu\)-D spectrograms. An adaptive technique is used to slice the overall captured signatures into half walking gait cycles. The experimental setup includes walking at different aspect angles with respect to the radar. The proposed approach is based on a similarity score methodology that derives a correlation between the walking signatures in different angles. 

\section{Micro-Motion Signatures}

\label{sec:micromotionsignatures}

\begin{figure}[tb]
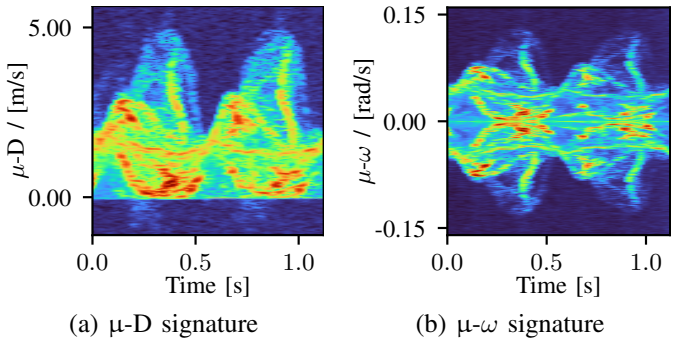

	\begin{minipage}[b]{.48\linewidth}
		\centering
		\centerline{\resizebox{1\linewidth}{!}{\input{fig/bigger/full_cycle_micro_Doppler.tex}}}
		\centerline{(a) \(\upmu\)-D signature}
	\end{minipage}
	\hfill
	\begin{minipage}[b]{.48\linewidth}
		\centering
		\centerline{\resizebox{1\linewidth}{!}{\input{fig/bigger/full_cycle_micro_omega.tex}}}
		\centerline{(b) \(\upmu\)-\(\omega\) signature}
	\end{minipage}
	\caption{Micro-motion signatures of a full gait cycle.}
	\label{fig:spectrograms}
	\vspace{-12 pt}
\end{figure}

	\par Each human has a walking signature that is defined mainly by the limbs' periodic swinging behavior within the gait cycles \cite{boulic1990global}. Such swinging motions induce angular displacement and velocity. A technique to estimate a walking human's tangential angular velocity is presented in \cite{nanzer2010millimeter}. A direct relation between the interferometric frequency and angular velocity is derived without relying on an AoA estimation. That algorithm was proposed to circumvent the limited AoA resolution in interferometric radar back then. However, this is not considered a limitation anymore due to the current high availability of MIMO modules in terms of cost, size-efficiency, and detection capability, e.g., range and AoA resolutions. The estimated velocity is proven to be complementary to the radial velocity \cite{nanzer2014dual} and reflects the micro-motion behavior in the tangential direction with respect to the radar \cite{nanzer2017microwave,nanzer2016micro,liang2020enhanced}. The main goal of that research was to increase the radar's FOV and to enhance the velocity estimation capabilities. Therefore, widely-spaced receiving antennas are required \cite{nanzer2010millimeter,merlo2021multiple}, which comes at the expense of AoA estimation accuracy. However, the latter is of high importance for surveillance.

	\par Our proposed technique for angular velocity estimation is different. It is based on a MIMO radar with frequency modulated continuous wave (FMCW) transmission protocol to estimate the micro-angular (\(\upmu\)-\(\omega\)) spectrograms through monitoring the rate of change of the received AoA profiles with respect to time. To the best of our knowledge, this has not been implemented in any previous study, and the \(\upmu\)-\(\omega\) signatures have not been deployed in any activity recognition use-cases yet. The derived \(\upmu\)-\(\omega\) signatures are expected to enhance the identification accuracy when combined with the commonly used \(\upmu\)-D signatures.

	\par The utilized radar sensor can capture information to formulate range-AoA maps for the sagittal plane \cite{geibig2016compact}. The range is estimated by evaluating the transmitted signal's round-trip time, where the range resolution is dependant on the transmission frequency bandwidth. The AoA estimation is based on the beamforming capabilities of multiple receiving (Rx) antennas. As described in \cite{milligan2005modern}, the AoA (\(\theta\)) and \unit[3]{dB} angular resolution (\(\theta_{res}\)) are:
	\begin{equation}
	\theta_{res} = \frac{1.78}{N_{Rx}}, \hspace{7 mm}\theta = \arcsin{\frac{\lambda\Delta\phi}{2\pi{d_a}}}
	\label{eqn:thres}	
	\end{equation}
	
	\par where \(N_{Rx}\) is the number of Rx antennas, \(d_a\) is the spacing between adjacent antennas, and \(\Delta\phi\) is the phase difference between two consecutive Rx antennas. Since \(\theta_{res}\propto N_{Rx}\), our algorithm deploys both available transmitting antennas to simulate a virtual antenna array with \(2\times N_{Rx}\) \cite{pirkani2019implementation}. A range-AoA map is formulated within each transmission chirp of duration (\(T_{ch}\)) and is then accumulated through time. To formulate spectrograms that reflect the micro-motion behavior, a short-time Fourier transform (STFT) is used for the estimation of the rate of change over time. Applying it to the AoA and range separately yields both spectrograms, namely \(\upmu\)-\(\omega\) and \(\upmu\)-D. The formulated \(\upmu\)-\(\omega\) spectrogram describes the elevation angular velocity, since the human limbs acquire angular displacements in the elevation plan while walking. From the resulting \(\upmu\)-D spectrogram the velocity bins within \(\pm\)\unit[0.02]{m/s} and beyond \(\pm\)\unit[6.0]{m/s} are discarded, effectively diminishing the effect of static objects and multi-path reflections for better training results \cite{vandersmissen2018indoor}. Both formulated spectrograms are synchronized in time. Fig.~\ref{fig:spectrograms} visualizes a full gait cycle of (a) the \(\upmu\)-D and (b) the \(\upmu\)-\(\omega\) spectrogram.
	
	\begin{figure}[tb]
		\begin{minipage}[b]{1\linewidth}
			\centering
			\centerline{\resizebox{1\linewidth}{!}{\input{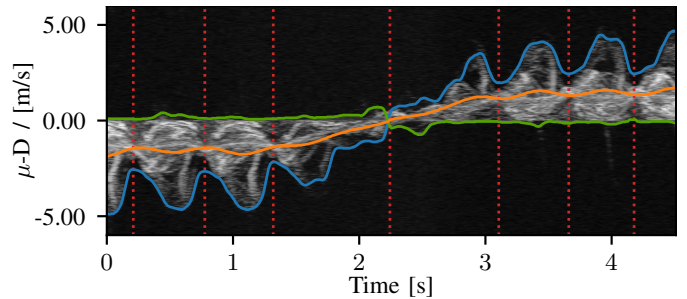}}}
		\end{minipage}
		\caption{\(\upmu\)-D signature at LOS with detected envelopes (blue and green), CoG (orange) and slicing positions (red).}
		\label{fig:envelopes}
		\vspace{-12 pt}
	\end{figure}
	
	\begin{figure*}[tb]
		\begin{minipage}[b]{.32\linewidth}
			\centering
			\centerline{\input{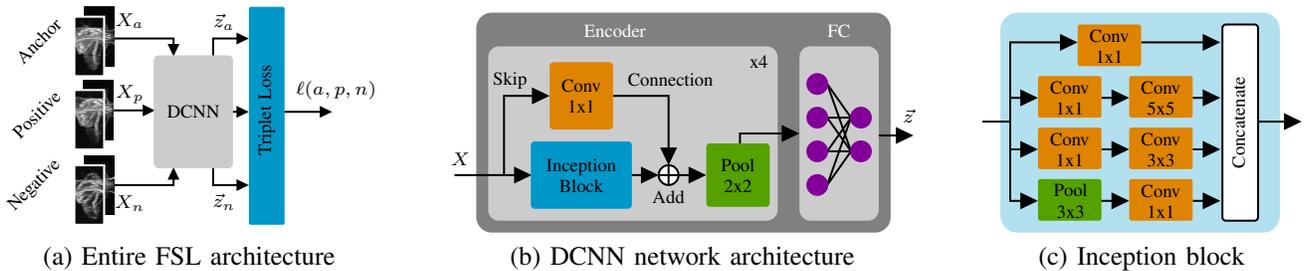}}
			\centerline{(a) Entire FSL architecture}
		\end{minipage}
		\hfill
		\begin{minipage}[b]{.37\linewidth}
			\centering
			\centerline{\tikzset{every picture/.style={line width=0.75pt}} 

\begin{tikzpicture}[x=0.75pt,y=0.75pt,yscale=-1,xscale=1]

\draw  [color={rgb, 255:red, 128; green, 128; blue, 128 }  ,draw opacity=0 ][fill={rgb, 255:red, 128; green, 128; blue, 128 }  ,fill opacity=1 ] (25,16.84) .. controls (25,12.51) and (28.51,9) .. (32.84,9) -- (226.16,9) .. controls (230.49,9) and (234,12.51) .. (234,16.84) -- (234,112.16) .. controls (234,116.49) and (230.49,120) .. (226.16,120) -- (32.84,120) .. controls (28.51,120) and (25,116.49) .. (25,112.16) -- cycle ;
\draw  [color={rgb, 255:red, 204; green, 204; blue, 204 }  ,draw opacity=0 ][fill={rgb, 255:red, 204; green, 204; blue, 204 }  ,fill opacity=1 ] (31,31.53) .. controls (31,28.48) and (33.48,26) .. (36.53,26) -- (171.47,26) .. controls (174.52,26) and (177,28.48) .. (177,31.53) -- (177,108.47) .. controls (177,111.52) and (174.52,114) .. (171.47,114) -- (36.53,114) .. controls (33.48,114) and (31,111.52) .. (31,108.47) -- cycle ;
\draw  [fill={rgb, 255:red, 255; green, 255; blue, 255 }  ,fill opacity=1 ] (117,91) .. controls (117,88.24) and (119.24,86) .. (122,86) .. controls (124.76,86) and (127,88.24) .. (127,91) .. controls (127,93.76) and (124.76,96) .. (122,96) .. controls (119.24,96) and (117,93.76) .. (117,91) -- cycle ; \draw   (117,91) -- (127,91) ; \draw   (122,86) -- (122,96) ;
\draw    (103,91) -- (114,91) ;
\draw [shift={(117,91)}, rotate = 180] [fill={rgb, 255:red, 0; green, 0; blue, 0 }  ][line width=0.08]  [draw opacity=0] (8.93,-4.29) -- (0,0) -- (8.93,4.29) -- cycle    ;
\draw    (127,91) -- (138,91) ;
\draw [shift={(141,91)}, rotate = 180] [fill={rgb, 255:red, 0; green, 0; blue, 0 }  ][line width=0.08]  [draw opacity=0] (8.93,-4.29) -- (0,0) -- (8.93,4.29) -- cycle    ;
\draw    (14,91) -- (50,91) ;
\draw [shift={(53,91)}, rotate = 180] [fill={rgb, 255:red, 0; green, 0; blue, 0 }  ][line width=0.08]  [draw opacity=0] (8.93,-4.29) -- (0,0) -- (8.93,4.29) -- cycle    ;
\draw    (39,91) -- (39,51) -- (59,51) ;
\draw [shift={(62,51)}, rotate = 180] [fill={rgb, 255:red, 0; green, 0; blue, 0 }  ][line width=0.08]  [draw opacity=0] (8.93,-4.29) -- (0,0) -- (8.93,4.29) -- cycle    ;
\draw    (94,51) -- (122,51) -- (122,83) ;
\draw [shift={(122,86)}, rotate = 270] [fill={rgb, 255:red, 0; green, 0; blue, 0 }  ][line width=0.08]  [draw opacity=0] (8.93,-4.29) -- (0,0) -- (8.93,4.29) -- cycle    ;
\draw  [color={rgb, 255:red, 204; green, 204; blue, 204 }  ,draw opacity=0 ][fill={rgb, 255:red, 204; green, 204; blue, 204 }  ,fill opacity=1 ] (188,31.53) .. controls (188,28.48) and (190.48,26) .. (193.53,26) -- (222.47,26) .. controls (225.52,26) and (228,28.48) .. (228,31.53) -- (228,108.47) .. controls (228,111.52) and (225.52,114) .. (222.47,114) -- (193.53,114) .. controls (190.48,114) and (188,111.52) .. (188,108.47) -- cycle ;
\draw    (202,45) -- (214,62) ;
\draw    (202,45) -- (214,79) ;
\draw    (202,62) -- (214,62) ;
\draw    (202,62) -- (214,79) ;
\draw    (202,96) -- (214,79) ;
\draw    (202,79) -- (211.04,79) -- (214,79) ;
\draw    (202,79) -- (214,62) ;
\draw    (202,96) -- (207.99,79.04) -- (214,62) ;
\draw  [color={rgb, 255:red, 130; green, 0; blue, 153 }  ,draw opacity=1 ][fill={rgb, 255:red, 130; green, 0; blue, 153 }  ,fill opacity=1 ] (192,96) .. controls (192,93.24) and (194.24,91) .. (197,91) .. controls (199.76,91) and (202,93.24) .. (202,96) .. controls (202,98.76) and (199.76,101) .. (197,101) .. controls (194.24,101) and (192,98.76) .. (192,96) -- cycle ;
\draw  [color={rgb, 255:red, 130; green, 0; blue, 153 }  ,draw opacity=1 ][fill={rgb, 255:red, 130; green, 0; blue, 153 }  ,fill opacity=1 ] (192,79) .. controls (192,76.24) and (194.24,74) .. (197,74) .. controls (199.76,74) and (202,76.24) .. (202,79) .. controls (202,81.76) and (199.76,84) .. (197,84) .. controls (194.24,84) and (192,81.76) .. (192,79) -- cycle ;
\draw  [color={rgb, 255:red, 130; green, 0; blue, 153 }  ,draw opacity=1 ][fill={rgb, 255:red, 130; green, 0; blue, 153 }  ,fill opacity=1 ] (192,62) .. controls (192,59.24) and (194.24,57) .. (197,57) .. controls (199.76,57) and (202,59.24) .. (202,62) .. controls (202,64.76) and (199.76,67) .. (197,67) .. controls (194.24,67) and (192,64.76) .. (192,62) -- cycle ;
\draw  [color={rgb, 255:red, 130; green, 0; blue, 153 }  ,draw opacity=1 ][fill={rgb, 255:red, 130; green, 0; blue, 153 }  ,fill opacity=1 ] (192,45) .. controls (192,42.24) and (194.24,40) .. (197,40) .. controls (199.76,40) and (202,42.24) .. (202,45) .. controls (202,47.76) and (199.76,50) .. (197,50) .. controls (194.24,50) and (192,47.76) .. (192,45) -- cycle ;
\draw  [color={rgb, 255:red, 130; green, 0; blue, 153 }  ,draw opacity=1 ][fill={rgb, 255:red, 130; green, 0; blue, 153 }  ,fill opacity=1 ] (214,62) .. controls (214,59.24) and (216.24,57) .. (219,57) .. controls (221.76,57) and (224,59.24) .. (224,62) .. controls (224,64.76) and (221.76,67) .. (219,67) .. controls (216.24,67) and (214,64.76) .. (214,62) -- cycle ;
\draw  [color={rgb, 255:red, 130; green, 0; blue, 153 }  ,draw opacity=1 ][fill={rgb, 255:red, 130; green, 0; blue, 153 }  ,fill opacity=1 ] (214,79) .. controls (214,76.24) and (216.24,74) .. (219,74) .. controls (221.76,74) and (224,76.24) .. (224,79) .. controls (224,81.76) and (221.76,84) .. (219,84) .. controls (216.24,84) and (214,81.76) .. (214,79) -- cycle ;
\draw    (158,75) -- (158,70) -- (185,70) ;
\draw [shift={(188,70)}, rotate = 180] [fill={rgb, 255:red, 0; green, 0; blue, 0 }  ][line width=0.08]  [draw opacity=0] (8.93,-4.29) -- (0,0) -- (8.93,4.29) -- cycle    ;
\draw    (228,71) -- (242.5,71) ;
\draw [shift={(245.5,71)}, rotate = 180] [fill={rgb, 255:red, 0; green, 0; blue, 0 }  ][line width=0.08]  [draw opacity=0] (8.93,-4.29) -- (0,0) -- (8.93,4.29) -- cycle    ;
\draw  [color={rgb, 255:red, 0; green, 0; blue, 0 }  ,draw opacity=0 ][fill={rgb, 255:red, 84; green, 161; blue, 0 }  ,fill opacity=1 ] (141,76.88) .. controls (141,75.84) and (141.84,75) .. (142.88,75) -- (171.13,75) .. controls (172.16,75) and (173,75.84) .. (173,76.88) -- (173,105.13) .. controls (173,106.16) and (172.16,107) .. (171.13,107) -- (142.88,107) .. controls (141.84,107) and (141,106.16) .. (141,105.13) -- cycle ;

\draw  [color={rgb, 255:red, 0; green, 0; blue, 0 }  ,draw opacity=0 ][fill={rgb, 255:red, 223; green, 132; blue, 0 }  ,fill opacity=1 ]  (62,36) .. controls (62,34.9) and (62.9,34) .. (64,34) -- (92,34) .. controls (93.1,34) and (94,34.9) .. (94,36) -- (94,66) .. controls (94,67.1) and (93.1,68) .. (92,68) -- (64,68) .. controls (62.9,68) and (62,67.1) .. (62,66) -- cycle  ;
\draw (78,51) node  [font=\scriptsize] [align=left] {\begin{minipage}[lt]{19.39pt}\setlength\topsep{0pt}
\begin{center}
Conv\\1x1
\end{center}

\end{minipage}};
\draw  [color={rgb, 255:red, 0; green, 0; blue, 0 }  ,draw opacity=0 ][fill={rgb, 255:red, 0; green, 152; blue, 201 }  ,fill opacity=1 ]  (52.5,76) .. controls (52.5,74.9) and (53.4,74) .. (54.5,74) -- (101.5,74) .. controls (102.6,74) and (103.5,74.9) .. (103.5,76) -- (103.5,106) .. controls (103.5,107.1) and (102.6,108) .. (101.5,108) -- (54.5,108) .. controls (53.4,108) and (52.5,107.1) .. (52.5,106) -- cycle  ;
\draw (78,91) node  [font=\scriptsize] [align=left] {\begin{minipage}[lt]{31.69pt}\setlength\topsep{0pt}
\begin{center}
Inception\\Block
\end{center}

\end{minipage}};
\draw (157,91) node  [font=\scriptsize] [align=left] {\begin{minipage}[lt]{17.01pt}\setlength\topsep{0pt}
\begin{center}
Pool\\2x2
\end{center}

\end{minipage}};
\draw (95,18) node  [font=\scriptsize,color={rgb, 255:red, 255; green, 255; blue, 255 }  ,opacity=1 ] [align=left] {Encoder};
\draw (167.5,33.5) node  [font=\scriptsize] [align=left] {x4};
\draw (208,18) node  [font=\scriptsize,color={rgb, 255:red, 255; green, 255; blue, 255 }  ,opacity=1 ] [align=left] {FC};
\draw (243,61) node  [font=\scriptsize]  {$\vec{z}$};
\draw (17,83) node  [font=\scriptsize]  {$X$};
\draw (122,102) node  [font=\scriptsize] [align=left] {Add};
\draw (123,43) node  [font=\scriptsize] [align=left] {Connection};
\draw (42,43) node  [font=\scriptsize] [align=left] {Skip};

\end{tikzpicture}}
			\centerline{(b) DCNN network architecture}
		\end{minipage}
		\hfill
		\begin{minipage}[b]{.27\linewidth}
			\centering
			\centerline{\tikzset{every picture/.style={line width=0.75pt}} 

\begin{tikzpicture}[x=0.75pt,y=0.75pt,yscale=-1,xscale=1]

\draw  [color={rgb, 255:red, 0; green, 152; blue, 201 }  ,draw opacity=0 ][fill={rgb, 255:red, 0; green, 152; blue, 201 }  ,fill opacity=0.3 ] (19,18.66) .. controls (19,13.88) and (22.88,10) .. (27.66,10) -- (151.34,10) .. controls (156.12,10) and (160,13.88) .. (160,18.66) -- (160,112.34) .. controls (160,117.12) and (156.12,121) .. (151.34,121) -- (27.66,121) .. controls (22.88,121) and (19,117.12) .. (19,112.34) -- cycle ;
\draw  [color={rgb, 255:red, 0; green, 0; blue, 0 }  ,draw opacity=1 ][fill={rgb, 255:red, 255; green, 255; blue, 255 }  ,fill opacity=1 ] (131,17.8) .. controls (131,16.81) and (131.81,16) .. (132.8,16) -- (147.2,16) .. controls (148.19,16) and (149,16.81) .. (149,17.8) -- (149,113.2) .. controls (149,114.19) and (148.19,115) .. (147.2,115) -- (132.8,115) .. controls (131.81,115) and (131,114.19) .. (131,113.2) -- cycle ;
\draw    (49,25) -- (55,25) ;
\draw [shift={(58,25)}, rotate = 180] [fill={rgb, 255:red, 0; green, 0; blue, 0 }  ][line width=0.08]  [draw opacity=0] (8.93,-4.29) -- (0,0) -- (8.93,4.29) -- cycle    ;
\draw    (24,53) -- (35,53) ;
\draw [shift={(38,53)}, rotate = 180] [fill={rgb, 255:red, 0; green, 0; blue, 0 }  ][line width=0.08]  [draw opacity=0] (8.93,-4.29) -- (0,0) -- (8.93,4.29) -- cycle    ;
\draw    (24,79) -- (35,79) ;
\draw [shift={(38,79)}, rotate = 180] [fill={rgb, 255:red, 0; green, 0; blue, 0 }  ][line width=0.08]  [draw opacity=0] (8.93,-4.29) -- (0,0) -- (8.93,4.29) -- cycle    ;
\draw    (30,105) -- (35,105) ;
\draw [shift={(38,105)}, rotate = 180] [fill={rgb, 255:red, 0; green, 0; blue, 0 }  ][line width=0.08]  [draw opacity=0] (8.93,-4.29) -- (0,0) -- (8.93,4.29) -- cycle    ;
\draw    (49,25) -- (24,25) -- (24,105) -- (30,105) ;
\draw    (10,65) -- (24,65) ;
\draw    (70,53) -- (81,53) ;
\draw [shift={(84,53)}, rotate = 180] [fill={rgb, 255:red, 0; green, 0; blue, 0 }  ][line width=0.08]  [draw opacity=0] (8.93,-4.29) -- (0,0) -- (8.93,4.29) -- cycle    ;
\draw    (70,79) -- (81,79) ;
\draw [shift={(84,79)}, rotate = 180] [fill={rgb, 255:red, 0; green, 0; blue, 0 }  ][line width=0.08]  [draw opacity=0] (8.93,-4.29) -- (0,0) -- (8.93,4.29) -- cycle    ;
\draw    (69,105) -- (81,105) ;
\draw [shift={(84,105)}, rotate = 180] [fill={rgb, 255:red, 0; green, 0; blue, 0 }  ][line width=0.08]  [draw opacity=0] (8.93,-4.29) -- (0,0) -- (8.93,4.29) -- cycle    ;
\draw    (116,53) -- (127,53) ;
\draw [shift={(130,53)}, rotate = 180] [fill={rgb, 255:red, 0; green, 0; blue, 0 }  ][line width=0.08]  [draw opacity=0] (8.93,-4.29) -- (0,0) -- (8.93,4.29) -- cycle    ;
\draw    (116,105) -- (127,105) ;
\draw [shift={(130,105)}, rotate = 180] [fill={rgb, 255:red, 0; green, 0; blue, 0 }  ][line width=0.08]  [draw opacity=0] (8.93,-4.29) -- (0,0) -- (8.93,4.29) -- cycle    ;
\draw    (116,79) -- (127,79) ;
\draw [shift={(130,79)}, rotate = 180] [fill={rgb, 255:red, 0; green, 0; blue, 0 }  ][line width=0.08]  [draw opacity=0] (8.93,-4.29) -- (0,0) -- (8.93,4.29) -- cycle    ;
\draw    (90,25) -- (127,25) ;
\draw [shift={(130,25)}, rotate = 180] [fill={rgb, 255:red, 0; green, 0; blue, 0 }  ][line width=0.08]  [draw opacity=0] (8.93,-4.29) -- (0,0) -- (8.93,4.29) -- cycle    ;
\draw    (149,65) -- (168,65) ;
\draw [shift={(171,65)}, rotate = 180] [fill={rgb, 255:red, 0; green, 0; blue, 0 }  ][line width=0.08]  [draw opacity=0] (8.93,-4.29) -- (0,0) -- (8.93,4.29) -- cycle    ;
\draw  [color={rgb, 255:red, 0; green, 0; blue, 0 }  ,draw opacity=0 ][fill={rgb, 255:red, 84; green, 161; blue, 0 }  ,fill opacity=1 ] (38,95.23) .. controls (38,94.55) and (38.55,94) .. (39.23,94) -- (68.77,94) .. controls (69.45,94) and (70,94.55) .. (70,95.23) -- (70,113.77) .. controls (70,114.45) and (69.45,115) .. (68.77,115) -- (39.23,115) .. controls (38.55,115) and (38,114.45) .. (38,113.77) -- cycle ;
\draw  [color={rgb, 255:red, 0; green, 0; blue, 0 }  ,draw opacity=0 ][fill={rgb, 255:red, 223; green, 132; blue, 0 }  ,fill opacity=1 ] (58,17.23) .. controls (58,16.55) and (58.55,16) .. (59.23,16) -- (88.77,16) .. controls (89.45,16) and (90,16.55) .. (90,17.23) -- (90,35.77) .. controls (90,36.45) and (89.45,37) .. (88.77,37) -- (59.23,37) .. controls (58.55,37) and (58,36.45) .. (58,35.77) -- cycle ;
\draw  [color={rgb, 255:red, 0; green, 0; blue, 0 }  ,draw opacity=0 ][fill={rgb, 255:red, 223; green, 132; blue, 0 }  ,fill opacity=1 ] (38,43.23) .. controls (38,42.55) and (38.55,42) .. (39.23,42) -- (68.77,42) .. controls (69.45,42) and (70,42.55) .. (70,43.23) -- (70,61.77) .. controls (70,62.45) and (69.45,63) .. (68.77,63) -- (39.23,63) .. controls (38.55,63) and (38,62.45) .. (38,61.77) -- cycle ;
\draw  [color={rgb, 255:red, 0; green, 0; blue, 0 }  ,draw opacity=0 ][fill={rgb, 255:red, 223; green, 132; blue, 0 }  ,fill opacity=1 ] (38,69.23) .. controls (38,68.55) and (38.55,68) .. (39.23,68) -- (68.77,68) .. controls (69.45,68) and (70,68.55) .. (70,69.23) -- (70,87.77) .. controls (70,88.45) and (69.45,89) .. (68.77,89) -- (39.23,89) .. controls (38.55,89) and (38,88.45) .. (38,87.77) -- cycle ;
\draw  [color={rgb, 255:red, 0; green, 0; blue, 0 }  ,draw opacity=0 ][fill={rgb, 255:red, 223; green, 132; blue, 0 }  ,fill opacity=1 ] (84,43.23) .. controls (84,42.55) and (84.55,42) .. (85.23,42) -- (114.77,42) .. controls (115.45,42) and (116,42.55) .. (116,43.23) -- (116,61.77) .. controls (116,62.45) and (115.45,63) .. (114.77,63) -- (85.23,63) .. controls (84.55,63) and (84,62.45) .. (84,61.77) -- cycle ;
\draw  [color={rgb, 255:red, 0; green, 0; blue, 0 }  ,draw opacity=0 ][fill={rgb, 255:red, 223; green, 132; blue, 0 }  ,fill opacity=1 ] (84,69.23) .. controls (84,68.55) and (84.55,68) .. (85.23,68) -- (114.77,68) .. controls (115.45,68) and (116,68.55) .. (116,69.23) -- (116,87.77) .. controls (116,88.45) and (115.45,89) .. (114.77,89) -- (85.23,89) .. controls (84.55,89) and (84,88.45) .. (84,87.77) -- cycle ;
\draw  [color={rgb, 255:red, 0; green, 0; blue, 0 }  ,draw opacity=0 ][fill={rgb, 255:red, 223; green, 132; blue, 0 }  ,fill opacity=1 ] (84,95.23) .. controls (84,94.55) and (84.55,94) .. (85.23,94) -- (114.77,94) .. controls (115.45,94) and (116,94.55) .. (116,95.23) -- (116,113.77) .. controls (116,114.45) and (115.45,115) .. (114.77,115) -- (85.23,115) .. controls (84.55,115) and (84,114.45) .. (84,113.77) -- cycle ;

\draw (74,27) node  [font=\scriptsize] [align=left] {\begin{minipage}[lt]{19.39pt}\setlength\topsep{0pt}
\begin{center}
Conv\\1x1
\end{center}

\end{minipage}};
\draw (139.2,65.24) node  [font=\scriptsize,rotate=-270] [align=left] {Concatenate};
\draw (54.5,105) node  [font=\scriptsize] [align=left] {\begin{minipage}[lt]{17.01pt}\setlength\topsep{0pt}
\begin{center}
Pool\\3x3
\end{center}

\end{minipage}};
\draw (54,53) node  [font=\scriptsize] [align=left] {\begin{minipage}[lt]{19.39pt}\setlength\topsep{0pt}
\begin{center}
Conv\\1x1
\end{center}

\end{minipage}};
\draw (54,79) node  [font=\scriptsize] [align=left] {\begin{minipage}[lt]{19.39pt}\setlength\topsep{0pt}
\begin{center}
Conv\\1x1
\end{center}

\end{minipage}};
\draw (100,53) node  [font=\scriptsize] [align=left] {\begin{minipage}[lt]{19.39pt}\setlength\topsep{0pt}
\begin{center}
Conv\\5x5
\end{center}

\end{minipage}};
\draw (100,79) node  [font=\scriptsize] [align=left] {\begin{minipage}[lt]{19.39pt}\setlength\topsep{0pt}
\begin{center}
Conv\\3x3
\end{center}

\end{minipage}};
\draw (100,105) node  [font=\scriptsize] [align=left] {\begin{minipage}[lt]{19.39pt}\setlength\topsep{0pt}
\begin{center}
Conv\\1x1
\end{center}

\end{minipage}};

\end{tikzpicture}}
			\centerline{(c) Inception block}
		\end{minipage}
		\vspace{-2mm}
		\caption{Overview of the FSL architecture (a) with a detailed view on the DCNN network (b) and its inception block (c) \cite{pegoraro2020multiperson}.}
		\label{fig:dcnnarchitecture}
		\vspace{-12 pt}
	\end{figure*}

\section{Proposed Algorithm}
	\label{sec:proposedalgorithm}

\subsection{Gait Cycle Slicing}
\label{sec:gaitcycleslicing}
	
	\par The gait cycle slicing is based on the grayscale \(\upmu\)-D spectrogram, which is visualized in Fig.~\ref{fig:envelopes}(a). Its primary (blue) and secondary (green) envelopes are computed based on a binarized version of the spectrogram that has been cleaned up through morphological closing \cite{9093181}. The grayscale version is used to compute the column-wise center of gravity (CoG, orange). Based on its value, the relative direction of movement from the target to the radar sensor can be determined. This information is crucial as Doppler shifts switch their sign depending on the direction of movement. By combining the information gained from the envelope's shape and the CoG, slicing locations can accurately be set. The \(\upmu\)-\(\omega\) spectrogram is sliced with accordance to the \(\upmu\)-D spectrogram. The useful spectrogram information within each slice is extracted by fitting a rectangle to the local envelopes' extrema and discarding all data points that lie outside it. Finally, concatenations of the extracted cohesive grayscale \(\upmu\)-D and \(\upmu\)-\(\omega\) spectrograms are fed to the FSL network (Fig.~\ref{fig:dcnnarchitecture}(a)). Due to the adaptive slicing, it can be expected that less noise is introduced into the dataset, which funnels down to a more stable and overall better training result. Radar-based human-ID on exact half gait cycles has not been explored for free walking yet.

\subsection{Few-Shot Learning}
\label{sec:humanidentification}
	\par While common DL techniques require multitudes of labeled data to perform satisfactory, FSL strives to achieve the same but with limited labeled data only. FSL is based on the idea of metric learning, thus relies on a similarity score in between two or more given samples. This similarity score is computed on an input data's embedding space representation, which is learned to contain the most meaningful features only. In contrast to conventional learning techniques, these features do not resemble the input data's shape but focus on discriminative characteristics in between samples of different classes. Hence, an embedding space is obtained that maps intrinsic data points to close proximity while pushing extrinsic ones further away.
	
	\begin{table}[!b]
		\vspace{-2mm}	
		\setlength\arrayrulewidth{0.01pt}
		\def\arraystretch{0.9}
		\scriptsize
		\centering
		\caption{MIMO radar module parametrization.  \label{tab:param}\vspace{-2mm}}
		\resizebox{\columnwidth}{!}{
			\begin{tabular}{l r l r}
				\toprule
				\multicolumn{2}{c}{Radar Parametrization} & \multicolumn{2}{c}{Attributes} \\
				\midrule
				Carrier frequency ($f_o$)  & \unit[77]{GHz} & {} & {}\\
				Tx-Rx antennas  & 2-16 & $\theta_{res}$ & \unit[3.19]{$^\circ$}\\
				Bandwidth ($B$)  & \unit[0.25]{GHz} & {} & {}\\
				Chirp duration ($T_{ch}$)   & \unit[80]{$\mu s$} & $v_{res}$ & \unit[4.753]{cm/s}\\
				Samples per chirp ($N_S$)   & 112 & {} & {}\\
				Chirps per frame ($N_P$)   & 512 & {} & {}\\
				\bottomrule
			\end{tabular}
		}
	\end{table}

	\par The identification task is carried out by a DCNN that is based on residual and inception structures \cite{szegedy2017inception}. The residual part is formed by skip connections (Fig.~\ref{fig:dcnnarchitecture}(b)) that learn the residual representation of the input data by feeding identity mappings deeper into the network. This allows for DCNNs to be trained faster and yields a considerable gain in performance. The inception block (Fig.~\ref{fig:dcnnarchitecture}(c)) is based on the idea of going wider instead of deeper. It is composed of parallel branches that extract differently scaled features. By factoring the convolution into a series of two operations, cross-channel and spatial correlations can be assessed independently by a 1x1, 3x3 and 5x5 convolution, respectively. As a result, inception blocks allow the network to learn richer representations of the input data with fewer parameters and are less prone to overfitting. The loss function is represented by the triplet loss. It takes three samples, namely an anchor (\(a\)), a positive (\(p\)) and a negative (\(n\)) with their respective class membership \(c_a\), \(c_p\) and \(c_n\) where \(c_a = c_p\) while \(c_a\neq c_n\). The triplet loss is formulated as 
	\begin{equation}
	\ell(a,p,n) = \max[d(a,p)-d(a,n)+\delta,0]
	\end{equation}
	where \(d\) denotes the Euclidean distance measure and \(\delta\) a tuneable margin. The loss is minimized by pushing \(d(a,p)\rightarrow0\) and \(d(a,n)\rightarrow\:>d(a,p)+\delta\). Batch-hard triplets are mined online during the training process, and mini-batches are assured to be balanced over the course of the training. While the former guarantees educational triplets throughout, the latter ensures an unbiased training result. The triplet loss is expected to derive a correlation between the captured micro-motion signatures of different aspect angles.

\section{Experimental Setup}
	\label{sec:experimentalsetup}

\begin{figure}[tb]
	\begin{minipage}[b]{1\linewidth}
		\centering
		\centerline{\resizebox{0.9\linewidth}{!}{\input{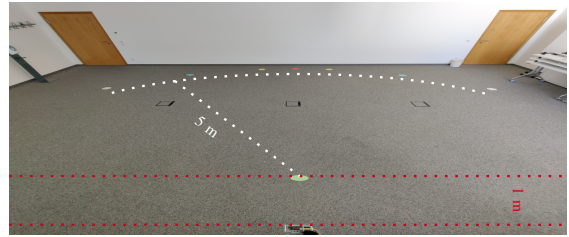}}}
	\end{minipage}
	\vspace{-3mm}
	\caption{Experimental setup for the non-constrained walking.}
	\label{fig:experimentalsetup}
	\vspace{-5mm}
\end{figure}
	
	\par The quality of the captured micro-motion signatures is affected when walking with an aspect angle to the radar line of sight (LOS) \cite{tahmoush2009angle}. Thus, walking at different aspect angles was considered to test the feasibility of the proposed triplet-loss approach. For the first dataset, 22 subjects walk according to their non-constrained, natural gait along predefined linear paths at the aspect angles \([\pm\unit[50]{^\circ},\pm\unit[30]{^\circ},\pm\unit[10]{^\circ},\unit[0]{^\circ}]\) with respect to LOS. Each path starts at \unit[1]{m} away from the radar and has a radial length of \unit[5]{m}, as shown in Fig.~\ref{fig:experimentalsetup}. The overall recording time per target is \unit[6]{min}. The main aspect is to achieve human-ID on a bigger FOV with better non-constrained walking. The entire dataset is split into \unit[70]{\%} for training and \unit[30]{\%} for validation. A separate test dataset of \unit[70]{s} per subject has also been collected. The utilized radar is parameterized as shown in Table \ref{tab:param}.
	
\section{Results and Discussion}
\label{sec:resultsanddiscussion}

\subsection{Spectrogram Constellations}
\label{sec:spectrogramconstellations}

\par Human-ID has mainly been based on the \(\upmu\)-D signature only. Deviating from this, the presented approach gives the opportunity to train on different spectrogram constellations. Three trainings are conducted on the non-constrained walking dataset. At first, trainings on both, the \(\upmu\)-D and \(\upmu\)-\(\omega\) are executed separately. Then, the feasibility of cohesive \(\upmu\)-D and \(\upmu\)-\(\omega\) concatenation is tested. All trainings include the entire 22 classes. The error rates are \unit[15.2]{\%}, \unit[11.2]{\%} and \unit[10.1]{\%}, respectively. Even though both of the spectrograms exhibit closely related information, the \(\upmu\)-\(\omega\) seems to capture more discriminative features than the well-explored \(\upmu\)-D one. The concatenated training shows a relative error reduction to the \(\upmu\)-D only case of \unit[33.6]{\%} and to the \(\upmu\)-\(\omega\) only case of \unit[9.8]{\%}. This proves the effectiveness of spectrogram concatenation which is used in all further experiments.

\subsection{Dataset Complexity}
\label{sec:datasetcomplexity}

	\par
\begin{figure}[tb]
	\begin{minipage}[b]{1\linewidth}
		\centering
		\centerline{\resizebox{1\linewidth}{!}{\input{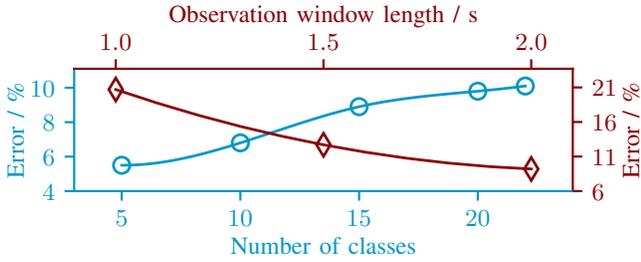}}}
	\end{minipage}
	\caption{Impact of the dataset complexity and the observation window length on the error rate of the FSL network.}
	\label{fig:complexityobservation}
	\vspace{-5mm}
\end{figure}

 As can be seen in prior research \cite{cao2018radar, qiao2020human, vandersmissen2018indoor}, the complexity of the dataset, which refers to the number of classes, has a severe impact on the accuracy of DCNN architectures that are based on conventional DL techniques. These challenges arise from their core working principle to immediately classify a query sample to a specific class. Since FSL learns a distance function over given samples in an embedding space, meaningful features can be extracted as long as naturally related data is presented. Thus, the classification performance is affected less by the dataset complexity. This behavior can be observed on the presented DCNN architecture when it is given concatenations of cohesive \(\upmu\)-D and \(\upmu\)-\(\omega\) spectrograms from the dataset of non-constrained movement of 5 to 22 different targets and is visualized in Fig. \ref{fig:complexityobservation} (blue). For five different classes in the test set, the presented approach yields an error of \unit[5.5]{\%} which increases to \unit[10.1]{\%} for 22 different classes. While the error rate still increases with an increasing dataset complexity, it is less distinct compared to previous studies.

\subsection{Dataset Size}
\label{sec:datasetsize}

\begin{table}[b]
	\vspace{-3mm}
	\setlength{\tabcolsep}{ 5 pt}
	\centering
	\caption{Impact of training dataset sizes on the error rate of the FSL network. The results are based on the test dataset.}
	\begin{tabular}{lcccc}
		\toprule
		& \multicolumn{4}{c}{Amount of data / min}\\
		\cmidrule(lr){2-5}
		& 1.05 & 2.10 & 3.15 & 4.20 \\
		\midrule
		Error rates [\%] & 24.9 & 11.3 & 10.4 & 10.1\\
		\bottomrule
	\end{tabular}
	\label{tab:datasetsize}
\end{table}

\par Since few-shot learning is mainly concerned about the limited data availability, the presented DCNN architecture is tested on different amounts of data. Recent studies use \(\approx\)\unit[20]{min} of data per subject on average. Accordingly, our collected data of non-constrained walking for \unit[6]{min} per target is considered limited. Nevertheless, to test the system's capability, the dataset size is further reduced. Table~\ref{tab:datasetsize} gives an overview on the sizes of the training dataset and the corresponding error rates. It can be reported that only \unit[1]{min} of training data would lead to a high error of \unit[24.9]{\%}. Nonetheless, by doubling this amount to only \unit[2]{min}, an error rate of \unit[11.3]{\%} is achieved, which wasn't heavily affected afterwards with bigger training data sizes of \unit[3.15,  4.20]{min}. This shows the merit of the proposed FSL approach on giving robust accuracy on unseen data even with very limited training duration.

\subsection{Observation Window}
\label{sec:spectrogramlength}
	
	\par To prove the feasibility of half gait cycle-based human-ID, a test series is executed where the observation window length is altered. To keep the effect of different dataset sizes to a minimum, they are kept about the same size throughout the experiments by changing the overlap between consecutive windows. Results are depicted in Fig. \ref{fig:complexityobservation} (red). Apart from the proposed half gait cycle approach, it can be noticed that error rates decrease with an increase in the observation window's length. This is due to the fact that fixing the observation window length cannot track the shape consistency of the extracted signatures. This non consistent signatures leads to a higher error rate for smaller observation window. The utilized \(\upmu\)-D spectrogram adaptive segmentation leads to a reliable extraction of the consistent shape of the captured gait cycles. As discussed in Sec.~\ref{sec:datasetcomplexity}, our proposed framework on half gait cycle ($\approx \unit[0.5]{s}$) yields a comparable classification error to the fixed observation window of \unit[2]{s}.

\subsection{Embedding Space}
\label{sec:embeddingspace}

\begin{figure}[tb]
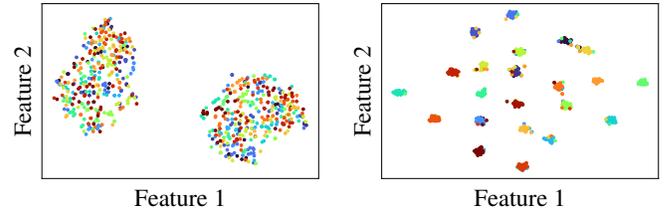

	\begin{minipage}[c]{.49\linewidth}
		\centering
		\centerline{\resizebox{1\linewidth}{!}{\input{fig/tsne_prior.pgf}}}
		\centerline{(a) t-SNE prior training}
	\end{minipage}
	\hfill
	\centering
	\begin{minipage}[c]{.49\linewidth}
		\centering
		\centerline{\resizebox{1\linewidth}{!}{\input{fig/tsne_after.pgf}}}
		\centerline{(b) t-SNE after training}
	\end{minipage}
	\caption{t-SNE visualizations of the embedding space.} 
	\label{fig:tsne}
	\vspace{-2.0mm}
\end{figure}
	
\par The FSL architecture's performance is highly dependent on the quality of its embedding space. The latter can be visualized through a t-distributed stochastic neighbor embedding (t-SNE) algorithm. It is depicted in Fig.~\ref{fig:tsne} for the proposed system on the test dataset (a) prior and (b) after the training process. The different colors represent the 22 different classes. Before training, two widely spread but well-separated clusters can be observed, where all classes are randomly mixed up. The two clusters are caused by the direction of relative movement towards or away from the radar. The randomly mixed up samples indicate that intra-class variances are high compared to inter-class variances. After the training has finished, individual classes are tightly clustered and well separated within the embedding space, reflecting the classification error of \unit[11.3]{\%}. 

\section{Conclusion}
\label{sec:conclusion}

\par This paper proposes a radar-based human-ID framework that is based on an inception-residual DCNN architecture. The need for large training datasets is overcome by leveraging FSL, and the triplet loss is successfully deployed to learn a discriminative embedding space on challenging data. Furthermore, the size of individual data samples is reduced to only half a gait cycle (\(\approx\)\unit[0.5]{s}). Each sample is the concatenation of time-synchronized \(\upmu\)-D and newly-formulated \(\upmu\)-\(\omega\) spectrograms, where the latter reflects the micro-motion behavior of the angular velocity in the elevation plane. The half walking gait cycles are segmented in an adaptive style. An error rate of \unit[11.3]{\%} is achieved on only \unit[2]{min} training data. Further research will include the implementation of open-set capabilities and running the framework on a real-time basis. Previous studies reported aspect angle $>\pm\unit[60]{^\circ}$ as the edge point between radial and tangential movements. Thus, this is considered as a limitation in using the micro-motion signatures for human-ID. In these cases, a 3D-point cloud radar can be beneficial to capture more descriptive features that can help in human-ID in static scenarios or for different motions.




\end{document}